\documentclass[apjl]{emulateapj}
\usepackage{amssymb}
\usepackage{epsfig}
\usepackage{url}

\begin{document}


\title{An HST/WFC3-UVIS View of the Starburst in the Cool Core of the Phoenix Cluster}


\author{Michael McDonald$^{1*\dagger}$, Bradford Benson$^2$, Sylvain Veilleux$^{3,4,5,6}$, Marshall W. Bautz$^1$, and Christian~L.~Reichardt$^7$ }
\altaffiltext{1}{Kavli Institute for Astrophysics and Space Research, MIT, Cambridge, MA 02139, USA}
\altaffiltext{2}{Kavli Institute for Cosmological Physics, University of Chicago, 5640 South Ellis Avenue, Chicago, IL 60637, USA}
\altaffiltext{3}{Department of Astronomy, University of Maryland, College Park, MD 20742, USA}
\altaffiltext{4}{Joint Space-Science Institute, University of Maryland, College Park, MD 20742, USA}
\altaffiltext{5}{Astroparticle Physics Laboratory, NASA Goddard Space Flight Center, Greenbelt, MD 20771, USA}
\altaffiltext{6}{Max-Planck-Institut f\"ur extraterrestrische Physik, Postfach 1312, D-85741 Garching, Germany}
\altaffiltext{7}{Department of Physics, University of California, Berkeley, CA 94720, USA}

\altaffiltext{*}{Email: mcdonald@space.mit.edu}
\altaffiltext{$\dagger$}{Hubble Fellow}


\begin{abstract}
We present \emph{Hubble Space Telescope Wide Field Camera 3} observations of the core of the Phoenix Cluster (SPT-CLJ2344-4243) in five broadband filters spanning rest-frame 1000--5500\AA. These observations reveal complex, filamentary blue emission, extending for $>$40~kpc from the brightest cluster galaxy. We observe an underlying, diffuse population of old stars, following an r$^{1/4}$ distribution, confirming that this system is somewhat relaxed. The spectral energy distribution in the inner part of the galaxy, as well as along the extended filaments, is a smooth continuum and is consistent with that of a star-forming galaxy, suggesting that the extended, filamentary emission is not due to the central AGN, either from a large-scale ionized outflow or scattered polarized UV emission, but rather a massive population of young stars. We estimate an extinction-corrected star formation rate of 798 $\pm$ 42 M$_{\odot}$ yr$^{-1}$, consistent with our earlier work based on low spatial resolution ultraviolet, optical, and infrared imaging. The lack of tidal features and multiple bulges, combine with the need for an exceptionally massive ($>10^{11}$ M$_{\odot}$) cold gas reservoir, suggest that this star formation is not the result of a merger of gas-rich galaxies.
Instead, we propose that the high X-ray cooling rate of $\sim$2700 M$_{\odot}$ yr$^{-1}$ is the origin of the cold gas reservoir. The combination of such a high cooling rate and the relatively weak radio source in the cluster core suggests that feedback has been unable to halt cooling in this system, leading to this tremendous burst of star formation.
\end{abstract}


\keywords{}


\section{Introduction}

In the cores of some galaxy clusters, the hot ($\gtrsim$10$^7$~K) intracluster medium (ICM) can reach high enough densities that the gas should cool by a combination of thermal bremsstrahlung and line cooling in less than a Hubble time, leading to cooling flows on the order of 100-1000 M$_{\odot}$ yr$^{-1}$ \citep[][]{fabian94}. Much effort has been devoted to detecting these cooling flows at lower temperatures using a variety of methods.  Intermediate temperature gas has been detected in a number of clusters via spectroscopy in the X-ray \citep{peterson01,peterson06}, UV \citep[e.g., \ion{O}{6};][]{oegerle01, bregman06}, optical \citep[e.g., H$\alpha$;][]{hu85,heckman89,crawford99,mcdonald10}, infrared \citep[e.g., H$_2$;][]{jaffe05,donahue11}, and millimeter \citep[e.g., CO;][]{edge01,edge02, salome03,mcdonald12b}.  However, even if this intermediate-temperature gas originated in the hot phase, the amount of it is orders of magnitude less than predicted by simple cooling flow models. Similarly, the star formation rate in the brightest cluster galaxy (BCG), whether measured in the UV/optical \citep{mcnamara89, koekemoer99, odea04,hicks10, mcdonald11b} or infrared \citep{odea08,rawle12,hoffer12}, only accounts for a small fraction ($\lesssim$10\%) of the predicted cooling flow. 
While there is some evidence that this intermediate-temperature gas and low-level star formation may represent a ``residual cooling flow'' \citep[e.g.,][]{peterson06,odea08, mcdonald10, hicks10, tremblay12}, the fact that the majority of the X-ray-inferred cooling is unaccounted for has served as prime evidence that some form of feedback is at work in cluster cores. The exact feedback mechanism that prevents the cooling catastrophe from happening is still not fully understood. The leading hypothesis of mechanical AGN-driven feedback is supported by the correlation between the total energy required to offset cooling and the amount of radio power required to inflate X-ray cavities \citep{rafferty06,fabian12,mcnamara12}.

\begin{figure*}[htb]
\centering
\includegraphics[width=0.95\textwidth]{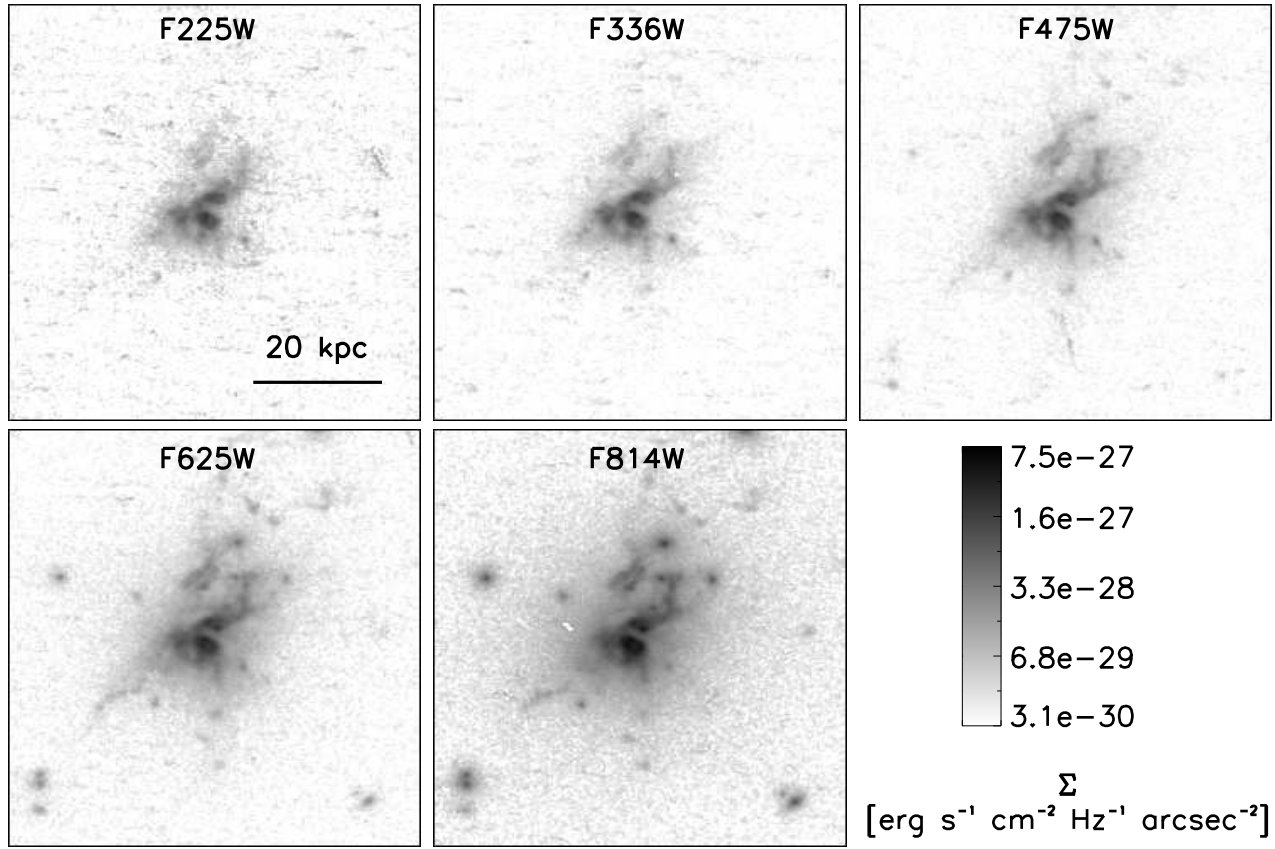}
\caption{HST-WFC3 images of the core of the Phoenix cluster, in 5 different optical bands (rest-frame 1000--5500\AA). These images highlight the complex morphology of the BCG. The presence of extended emission in all bands argues in favor of continuum emission from young stars over line emission from ionized gas. The full extent of these filaments ($\sim$40~kpc) is reminiscent of those observed in nearby cool core clusters such as Perseus and Abell~1795.}
\label{fig:3x2}
\end{figure*}

Recently, \cite{mcdonald12c} reported the unique properties of a galaxy cluster at $z=0.596$, SPT-CLJ2344-4243 (hereafter the Phoenix Cluster), which was initially discovered by the South Pole Telescope using the Sunyaev Zel'dovich effect \citep{williamson11}.
This cluster is among the most massive (M$_{200,Y_X}\sim2.5\times10^{15}$ M$_{\odot}$), X-ray luminous (L$_{2-10\rm{keV}}=8.2\times10^{45}$ erg s$^{-1}$), and strongest cooling (\.{M}~$\sim$~3820~M$_{\odot}$~yr$^{-1}$) clusters discovered to date. The central galaxy in the Phoenix cluster appears to be experiencing a 740 M$_{\odot}$ yr$^{-1}$ starburst. However, it also harbors a powerful AGN which makes it difficult to separate contributions to the ultraviolet, optical, and infrared emission from star-forming regions versus the AGN in low-resolution imaging. Further, the available ground-based data optical and space-based ultraviolet (UV) and infrared (IR) data presented in \cite{mcdonald12c} were unable to distinguish between in situ star formation or the late stages of a merger.

In this Letter, we present new \emph{Hubble Space Telescope} observations which improve significantly in depth and spatial resolution on the data presented in \cite{mcdonald12c}. In \S2 we describe our analysis of these data, after which we present our new, detailed view of the Phoenix cluster in \S3. In \S4 we discuss the possible interpretations of these data, including whether or not we can separate the cooling flow scenario from a pure AGN or merger scenario. We conclude with a summary of these results in \S5. Throughout this letter we assume H$_0$~=~70 km~s$^{-1}$~Mpc$^{-1}$, $\Omega_M$~=~0.27, and $\Omega_{\Lambda}$~=~0.73.


\section{HST Data}

To study in detail the purported starburst in the core of the Phoenix cluster, we obtained broadband imaging with the \emph{Hubble Space Telescope} \emph{Wide Field Camera 3} (HST WFC3) in five optical filters - F225W, F336W,  F475W, F625W, F814W - which span rest-frame wavelengths from $\sim$1000\AA\ to $\sim$5500\AA. These observations were carried out over 2 orbits of Director's Discretionary Time, with an average exposure time of $\sim$1800s per filter (PID \#13102, PI McDonald). 

In each filter, a series of two dithered exposures were obtained. The LA Cosmic\footnote{\url{http://www.astro.yale.edu/dokkum/lacosmic/}} \citep{lacosmic} software was run on individual exposures, generating accurate cosmic ray and bad pixel masks for each image. Pairs of exposures were then combined using the {\sc PyRAF} {\sc astrodrizzle} routine\footnote{\url{http://www.stsci.edu/hst/HST\_overview/drizzlepac}}, with the aforementioned masks preceding the standard cosmic ray detection in the {\sc multidrizzle} task. The final, cleaned images are presented in Figure \ref{fig:3x2}.

All optical and UV fluxes were corrected for Galactic extinction following \cite{cardelli89} using a Galactic reddening estimate of $E(B-V)=0.017$ towards the cluster center, from \cite{schlegel98}.

\section{Results}

\begin{figure*}
\centering
\includegraphics[width=0.99\textwidth]{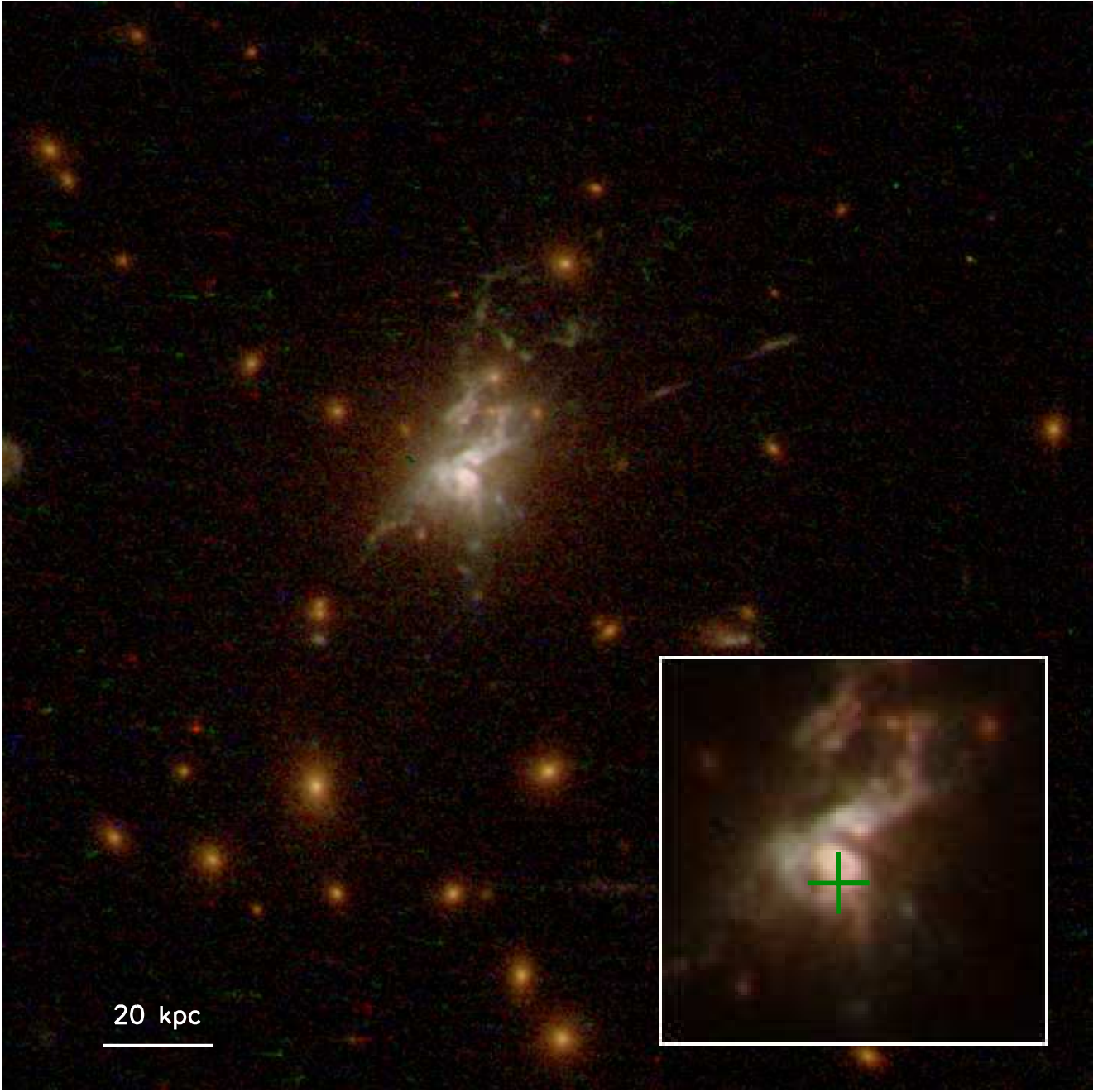}
\caption{Color image, combining the F475W, F625, and F814W bands, showing the young, filamentary, star-forming regions overlaid on the diffuse, old stellar component of the BCG. The most extended filaments in this complex system extend for $\sim$40~kpc to the north and northwest of the cluster center. The linear, radial feature to the northwest has colors consistent with the rest of the filaments, suggesting that it is neither a jet nor a background, lensed galaxy. In the inset we show the central $\sim$20~kpc, with the position (and uncertainty) of the X-ray point source in green. There appear to be gaps in the emission to the northwest and southeast of the peak, possibly due to strong dust lanes. }
\label{fig:rgb}
\end{figure*}

\begin{figure*}
\centering
\includegraphics[width=0.99\textwidth]{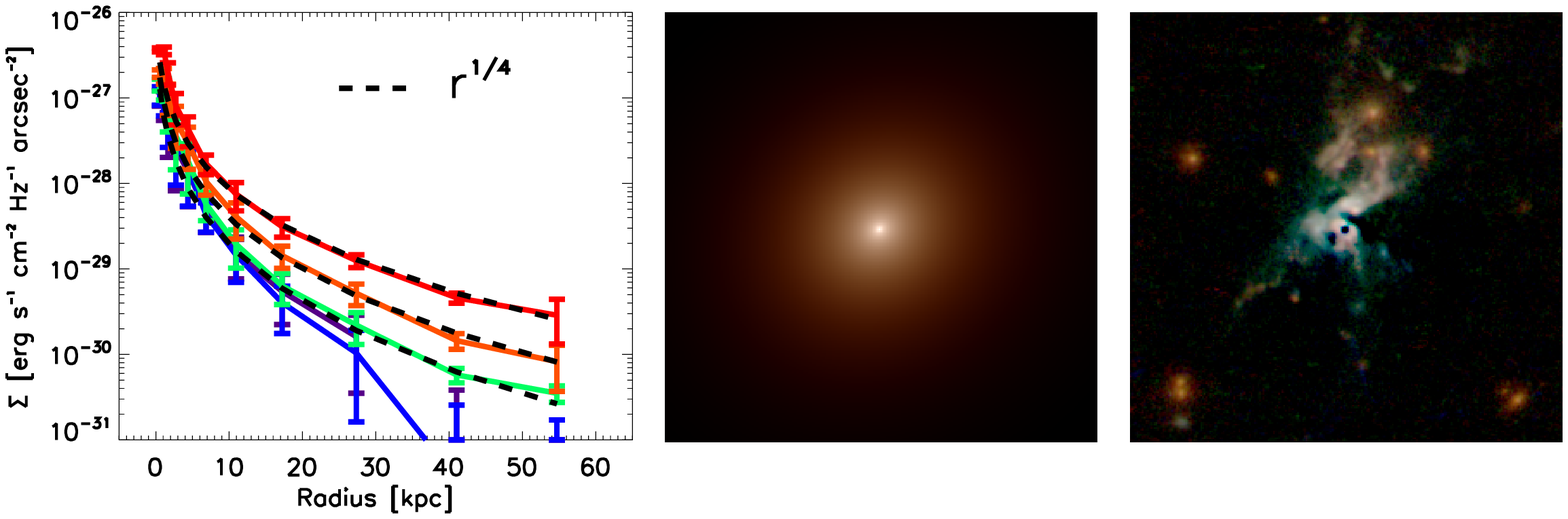}
\caption{Left: Surface brightness profiles of the BCG in all five bands (purple: F225W, blue: F336W, green: F475W, orange: F625W, red: F814W) . Each profile represents the average along four radial cuts with orientations chosen to avoid most of the extended, filamentary emission. The diffuse, red component of the BCG has a surface brightness distribution typical of an elliptical galaxy, well modeled by a $r^{1/4}$ fall-off in surface brightness \citep[dashed line;][]{devaucouleurs48}. Middle: 2-dimensional surface brightness model from GALFIT \citep{galfit}, with ellipticity and PA of 0.88 and -52$^{\circ}$, respectively. Right: Residual image generated by subtracting the r$^{1/4}$ model from each of the three reddest bands. Both color images use the same color scaling as Figure \ref{fig:rgb}. This figure highlights the complex morphology of the star-forming filaments superimposed on the smooth, old stellar population in the BCG.}
\label{fig:sbprof}
\end{figure*}

In Figure \ref{fig:3x2} we show the newly-acquired far-UV through optical HST images of the core of the Phoenix cluster, which provide a detailed picture of this system. These images show significant, extended filamentary emission at all wavelengths from $\sim$1000--5500\AA, overlaid on a relatively smooth population of older, red stars. The most extended pair of filaments to the north of the BCG are $\sim$6$^{\prime\prime}$ (40~kpc) in length, similar to the most extended filaments seen in Abell~1795 \citep{mcdonald09},  and the Perseus cluster \citep{conselice01,fabian08}. We measure a total rest-frame far-UV flux density of f$_{F225W}$ = 1.26$\times$10$^{-27}$ erg s$^{-1}$ cm$^{-2}$ Hz$^{-1}$, consistent with the GALEX-derived flux presented in \cite{mcdonald12c}. 

The fact that such complex, filamentary morphology is present in all five filters suggests that the BCG is forming stars at a prodigious rate. In the wavelength range covered, there may be contributing emission from the \ion{C}{4} $\lambda$1549 (F225W), [\ion{O}{2}] (F625W), and [\ion{O}{3}] and H$\beta$ (F814W) lines. However, the F336W and F475W bands, which have similar surface brightnesses to the other three bands, should be relatively free from emission lines, suggesting that young stars, not ionized gas, is the dominant source of the observed flux in Figure \ref{fig:3x2}.

In Figure \ref{fig:rgb} we show a three-color (F475W, F625W, F814W) image of the cluster core. This figure shows a clear difference in the stellar populations between the young (blue) filaments and the underlying, smoothly-distributed, old (red) stars. The peak of the emission in all bands is coincident (within the positional uncertainties) with the X-ray point source. To the northwest and southeast of the emission peak are dark lanes, most likely due to obscuration by dust. Overall, the color of the filamentary emission appears roughly constant with radius, and is reminiscent of a young, star-forming galaxy. We see no evidence for multiple bulges or tidal features, both of which would indicate that this system is the result of a recent merger of gas-rich galaxies.

Figure \ref{fig:sbprof} shows the multi-band surface brightness profiles of the BCG (left panel), which have been computed along radial cuts at four different angles (90$^{\circ}$, 120$^{\circ}$, 180$^{\circ}$, 210$^{\circ}$), chosen to avoid the blue, filamentary emission. The radial surface brightness profile follows an r$^{1/4}$ profile, which is typical of relaxed, early-type galaxies \citep{devaucouleurs48}. Such r$^{1/4}$ surface brightness distributions are also common in the final stages (single-nucleus) of low-redshift ($z<0.3$) gas-rich mergers \citep[ULIRGs; e.g.,][]{veilleux02, veilleux06}. However, with a half-light radius of $\sim$17~kpc and a stellar mass of $\sim3\times10^{12}$~M$_{\odot}$ \citep{mcdonald12c}, this BCG is a factor of $\sim$4 larger in size \citep{veilleux02}, a factor of $\sim$60 higher in stellar mass \citep{veilleux06}, and resides in an environment $\sim$50--100 times richer \citep{zauderer07} than normal for $z<0.3$ ULIRGs. Projecting these 1-dimensional profiles back onto the sky, we can separate diffuse, giant elliptical emission (middle panel) from clumpy, star-forming emission (right panel). The lack of smooth, arcing tidal features and multiple bulges in the residual image (right panel) suggests that this complex, extended emission did not originate from a recent merger with one or more gas-rich disk galaxies. All of these factors argue that the Phoenix BCG is unlike a traditional ULIRG by any definition other than the high total infrared luminosity.

\begin{figure*}
\centering
\includegraphics[width=0.99\textwidth]{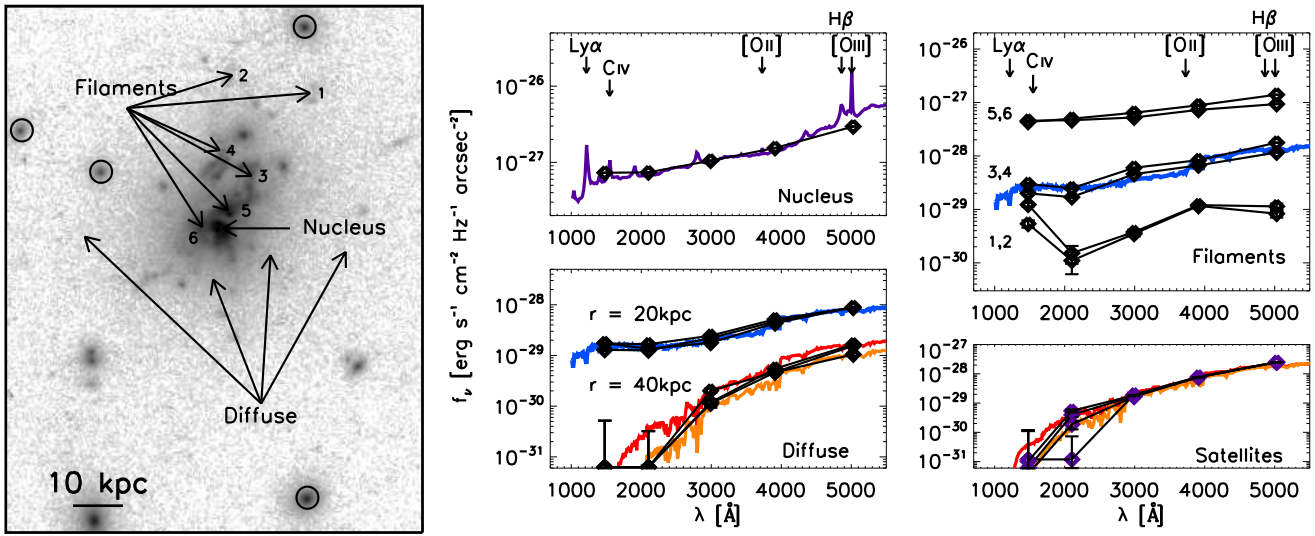}
\caption{Left: F814W image of the Phoenix BCG, with positions of nucleus, filaments, and diffuse emission indicated. Right: SEDs for the  regions indicated on the left. For comparison, we show the SEDs for four early-type satellite galaxies (marked by circles in the left panel)  which have been normalized to a single value of $f_{5000\AA}$. Orange and red spectra correspond to passive stellar populations of ages 2 and 5 Gyr, respectively, while blue and purple spectra correspond to a typical late-type spiral galaxy (Sdm) and dusty type-2 QSO, respectively \citep{polletta07}. All model spectra are normalized by eye and are meant to demonstrate the qualitative agreement in spectral shape. The faintest (labels 1,2) filaments show evidence for nebular emission, with peaks in the F225W and F625W bands, consistent with \ion{C}{4} and [\ion{O}{2}] emission. This figure suggests that, at large radii the diffuse stellar populations are typical for an early-type galaxy, while at small radii both the diffuse and filamentary emission result from significantly younger stellar populations. 
}
\label{fig:sed}
\end{figure*}

In Figure \ref{fig:sed} we provide the spectral energy distribution (SED) in several representative regions around the BCG. The diffuse emission at large and small radii indicate a significant positive age gradient in the diffuse population. At large (r $\sim$40~kpc) radii, the diffuse emission is qualitatively consistent with a 2--5 Gyr old elliptical galaxy, while at smaller radii ($\sim$20~kpc) the diffuse stellar populations appear to be much younger (bluer), similar in color to a star-forming spiral galaxy. The extended, morphologically-complex filaments, after subtraction of the diffuse stellar component, show an excess of UV emission at all radii. The SED of the brightest filaments appear remarkably similar to the diffuse component in the central region, suggesting that these stars are being mixed on short timescales. In the faintest, most extended filaments, there is a substantial excess of emission in the F225W and F625W filters, at the location of  the redshifted \ion{C}{4} $\lambda$1549 and [\ion{O}{2}] lines, respectively, suggesting that these filaments may also contain warm ($>10^4$K), ionized gas -- a scenario supported by the extended emission lines reported in \cite{mcdonald12c}. The overall flatness of the SED in the UV-bright regions is exactly what one would expect for a mix of young stars and warm, ionized gas, given the width of the broadband filters. At the peak of the optical emission, coincident with the X-ray source, the SED is qualitatively well matched by a dusty type-2 QSO \citep{polletta07}, which is consistent with our X-ray observations of a highly-reddened hard X-ray point source. 

The combination of Figures \ref{fig:3x2}--\ref{fig:sed} paint a picture of an old, giant elliptical galaxy that is experiencing a resurgence of star formation. Below we re-evaluate the SFR in this system and describe various scenarios to explain this star-forming activity, building on the discussion of \cite{mcdonald12c}.


\section{Discussion}
The deep, high spatial resolution HST UV and optical imaging presented in \S3 have revealed an exceptionally complex system. Below we utilize this improved spatial resolution to estimate a new, UV-derived SFR for the BCG in the Phoenix cluster, followed by a discussion of three possible origins for the extended, filamentary UV emission.

\subsection{A Revised Estimate of the SFR}
The SFR reported in \cite{mcdonald12c}, while utilizing an array of multi-wavelength data, necessarily required multiple assumptions to remove AGN contamination. With the addition of high-resolution HST UV imaging, we avoid such assumptions, leaving only the (typical) assumptions of dust extinctions and star formation laws.

\begin{figure}
\centering
\includegraphics[width=0.49\textwidth]{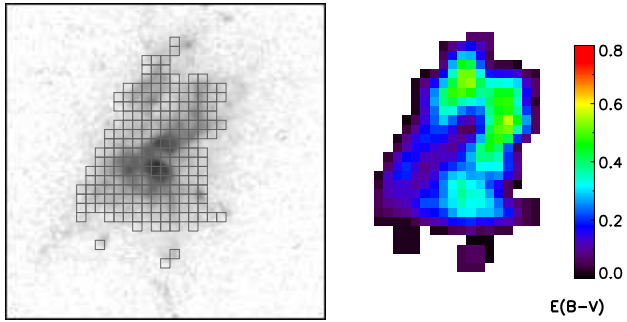}
\caption{Left: Image in the F475W band, with 0.2$^{\prime\prime}$ $\times$ 0.2$^{\prime\prime}$ bins overlaid in regions with $>5\sigma$ detections in both the F336W and F475W bands. Right: Smoothed reddening map, derived assuming a flat UV SED in the absence of reddening \citep{kennicutt98}. This map allows us to perform \emph{local} reddening corrections on the UV images, yielding a more tightly constrained estimate of the extinction-corrected UV luminosity.}
\label{fig:ebv}
\end{figure}

To estimate the reddening, we assume that the reddening-free SED is flat at 1500--3000\AA\ \citep{kennicutt98}. Since the F225W filter may be contaminated by \ion{C}{4} emission, we use the mean $f_{F336W}/f_{F475W}$ (rest frame $f_{210}/f_{298}$) flux ratio to derive a reddening correction, assuming $f_{F336W}/f_{F475W}=1$ for E(B-V) = 0. In Figure \ref{fig:ebv} we show the spatial distribution of the reddening, E(B-V), over regions with significant ($S/N>5$) UV flux. We note that the mean reddening in this map agrees well with the reddening presented in \cite{mcdonald12c} from Balmer line ratios (E(B-V) = 0.34). Using the \emph{local} reddening correction and uncertainty we can accurately correct the UV luminosity, resulting in a more confident estimate of L$_{2000}$ = 5.7 $\pm$ 0.3 $\times$ 10$^{30}$ erg s$^{-1}$ Hz$^{-1}$ and a UV-derived SFR of 798 $\pm$ 42 M$_{\odot}$ yr$^{-1}$ \citep{kennicutt98}.  Given the near-IR-estimated stellar mass of M$_* \sim 3\times10^{12}$ M$_{\odot}$ \citep{mcdonald12c}, this high SFR implies that in 100~Myr the central galaxy in the Phoenix cluster will form 2--3\% of its stellar mass.

We note that this estimate is consistent with the AGN-subtracted SFR quoted in \cite{mcdonald12c} of 739 $\pm$ 160 M$_{\odot}$ yr$^{-1}$ and with the empirical, extinction-implicit, method of \cite{rosa-gonzalez02}, which yields SFR$_{UV} = 1040^{+1290}_{-580}$ M$_{\odot}$ yr$^{-1}$.

\subsection{Possible Sources of Extended UV Emission}
\subsubsection{AGN-driven Outflow and Scattered Light}

Inspired by the high far-infrared luminosity, coupled with the hard X-ray and radio source, a viable explanation for this system is a dusty AGN driving a large-scale outflow. 
While there is undoubtedly a powerful AGN in the core of the Phoenix cluster, the new HST data suggest that the substantial UV luminosity can not be fully attributed to a point-like AGN. Only a small fraction ($<$10\%) of the total UV luminosity originates from a central point source, with the majority originating from extended, filamentary regions. Furthermore, the high relative fluxes in the F336W and F475W bands, which should be free from line emission, suggest that the majority of the UV/optical flux in these complex filaments is continuum emission. We also find a general lack of UV/blue emission along the minor axis of the central galaxy, contrary to what is typically observed in wide-angle AGN-driven outflows \citep{veilleux05}. Thus, we argue that the filamentary UV emission does not result from an outflow.

 In IRAS~09104+4109 \citep[e.g.,][]{osullivan12}, along with other powerful radio galaxies, much of the extended UV continuum is due to scattered, polarized light from a heavily-obscurred QSO. However, the UV continuum in IRAS~09104+4109 is $\sim$20 times fainter than that presented here for the Phoenix cluster \citep{hines99}. Considering the fact that IRAS~09104+4109 is already an extreme system, it seems unlikely that this additional factor of 20 can be accounted for with the same model.

\subsubsection{Gas Rich Merger(s)}
The majority of known ultraluminous infrared galaxies (ULIRGs) appear to be late-stage mergers of gas-rich galaxies \citep[$\sim$95\%;][]{veilleux02}, which begs the question: Is the extreme star formation in the core of the Phoenix cluster fueled by a gas-rich merger? Assuming a relation between molecular gas depletion time and specific star formation rate \citep{saintonge11}, we estimate an H$_2$ depletion time of 2--12$\times$10$^8$~yr, which is consistent with those measured in two nearby cooling flows: A1068 \citep[4$\times10^8$~yr;][]{edge01} and A1835 \citep[5$\times10^8$~yr;][]{edge01} . 
Assuming a 1-to-1 correspondence between the total mass of stars formed and the mass of the cold gas reservoir, this timescale implies a molecular gas mass of 1.5--9.5$\times10^{11}$ M$_{\odot}$, consistent with more recent work by \cite{combes12}, yielding 1.0--5.0$\times10^{11}$ M$_{\odot}$. 
This cold gas mass is significantly higher than that measured for gas-rich galaxies in the Virgo cluster \citep[e.g.,][]{corbelli12}. Further, the process of increasing the cold gas content by compressing atomic gas during a merger would be highly inefficient in the core of the Phoenix cluster due to ram pressure stripping of the HI disk during the initial galaxy infall - a process observed in the Virgo cluster, which has an ICM density $\sim$10 times lower than Phoenix.

\subsubsection{Cooling of the Intracluster Medium}
Our preferred explanation in \cite{mcdonald12c} was that the star formation in the cluster core is being fueled by gas cooling out of the ICM. This remains the most plausible avenue for such a large amount of cold gas to reach the core of the cluster, and is supported by the exceptionally bright X-ray cool core and relatively weak radio source. Such an imbalance between cooling and feedback could lead to rapid cooling of the ICM, fueling bursts of star formation. 

Following \cite{white97}, we estimate from the X-ray data the mass deposition rate of the cooling flow, combining the X-ray cooling luminosity with the gravitational potential of the cluster in order to correct for gravitational work done as the gas falls towards the cluster center. We obtain an ICM cooling rate of \.{M}~=~2700~$\pm$~700~M$_{\odot}$ yr$^{-1}$ which is enough to fuel a 798 M$_{\odot}$ yr$^{-1}$ starburst, assuming the feedback mechanism that prevents star formation in nearby clusters is operating less efficiently in Phoenix. In the future, high resolution X-ray spectroscopy of the cool core \citep[e.g.,][]{peterson06} will provide firm estimates of the ICM cooling rate down to low temperatures, revealing whether or not this scenario is a plausible fuel source for this massive starburst.

\section{Summary and Conclusions}

We report new HST observations of the Phoenix cluster (SPT-CLJ2344-4243) with WFC3-UVIS in five filters covering rest-frame wavelengths 1000-5500\AA. The high spatial resolution of HST is able to separate bright UV emission from the AGN and the surrounding diffuse, extended emission, definitively confirming the presence of a starburst in the BCG. The morphology of this central galaxy is complex, with narrow filaments extending for $>$40~kpc, reminiscent of the nearby Perseus and Abell~1795 clusters. We argue that the majority of the observed UV emission is due to young stars, on the basis of the complex morphology and flat SED over the wavelength range 1000--5500\AA. We confirm the high SFR presented in \cite{mcdonald12c}, measuring an extinction-corrected, AGN-removed UV-derived SFR of 798 $\pm$ 42 M${\odot}$ yr$^{-1}$.  We find that merger driven scenarios would require an unreasonably large number of gas rich galaxies to supply the cold gas reservoir required to fuel the starburst, and conclude that the starburst is likely fueled by a massive cooling flow.

\section*{Acknowledgements} 
We thank the HST Director for graciously providing the data for this program. M. M. acknowledges support provided by NASA through a Hubble Fellowship grant from STScI. S.V. acknowledges support from a Senior NPP Award held at
NASA-GSFC and from the Humboldt Foundation to fund a long-term visit at MPE in 2012. BAB is supported by the National Science Foundation through grant ANT-0638937, with partial support provided by NSF grant PHY-1125897, the Kavli Foundation, and Chandra Award Number 13800883 issued by the CXC.


\end{document}